\documentclass[review]{elsarticle}

\usepackage{graphicx,amsmath}
\usepackage{hyperref}
\usepackage{xspace}
\usepackage{xcolor}
\usepackage{ulem}
\usepackage{placeins}

\newcommand{\fig}[1]{Fig.~\ref{#1}}

\newcommand{\tab}[1]{Table \ref{#1}}
\newcommand{\secref}[1]{Section~\ref{#1}}
\newcommand{\secrefs}[2]{Sections~\ref{#1} and \ref{#2}}

\newcommand{\errval}[2]{${#1} \pm {#2}$}

\newcommand{\GeV}{\ensuremath{\mathrm{\,Ge\kern -0.1em V}}\xspace}
\newcommand{\MeV}{\ensuremath{\mathrm{\,Me\kern -0.1em V}}\xspace}
\newcommand{\GeVMom}{\ensuremath{{\mathrm{\,Ge\kern -0.1em V\!/}c}}\xspace}
\newcommand{\MeVMass}{\ensuremath{{\mathrm{\,Me\kern -0.1em V\!/}c^2}}\xspace}

\begin{document}

\title{Test-beam studies of a small-scale TORCH time-of-flight demonstrator}


\author[2,5]{S.~Bhasin}
\author[3]{T.~Blake}
\author[5]{N.~Brook}
\author[8]{T.~Conneely}
\author[2]{D.~Cussans}
\author[6]{R.~Forty}
\author[6]{C.~Frei}
\author[4]{E.~P.~M.~Gabriel}
\author[1]{R.~Gao}
\author[3]{T.~Gershon}
\author[6]{T.~Gys}
\author[1]{T.~Hadavizadeh}
\author[1]{T.~H.~Hancock\corref{cor1}}\ead{thomas.hancock@physics.ox.ac.uk}
\author[1]{N.~Harnew}
\author[3]{M.~Kreps}
\author[8]{J.~Milnes}
\author[6]{D.~Piedigrossi}
\author[2]{J.~Rademacker}
\author[7]{M.~van~Dijk}

\address[1]{Denys Wilkinson Laboratory, University of Oxford, Keble Road, Oxford OX1 3RH, United Kingdom}
\address[2]{H.H. Wills Physics Laboratory, University of Bristol, Tyndall Avenue, Bristol BS8 1TL, United Kingdom}
\address[3]{Department of Physics, University of Warwick, Coventry, CV4 7AL, United Kingdom}
\address[4]{School of Physics and Astronomy, University of Edinburgh, James Clerk Maxwell Building, Edinburgh EH9 3FD, United Kingdom}
\address[5]{University of Bath, Claverton Down, Bath BA2 7AY, United Kingdom}
\address[6]{CERN, EP Department, CH-1211 Geneva 23, Switzerland}
\address[7]{CERN, EN Department, CH-1211 Geneva 23, Switzerland}
\address[8]{Photek Ltd., 26 Castleham Road, St Leonards on Sea, East Sussex, TN389 NS, United Kingdom}

\cortext[cor1]{Corresponding author}

\begin{abstract}

TORCH is a time-of-flight detector designed to perform particle identification over the momentum range $2-10\,\rm{GeV/\textit{c}}$ for a 10\,m flight path. The detector exploits  prompt Cherenkov light produced by charged particles traversing a quartz plate of $10\,\rm{mm}$ thickness. Photons are then trapped by total internal reflection and directed onto a detector plane instrumented with customised  position-sensitive Micro-Channel Plate Photo-Multiplier Tube (MCP-PMT) detectors. A single-photon timing resolution of $70\,\rm{ps}$ is targeted to achieve the desired separation of pions and kaons, with an expectation of  around 30 detected photons per track. Studies of the performance of a small-scale TORCH demonstrator with a radiator of dimensions $120 \times 350 \times 10\,\rm{mm^3}$ have been performed in two test-beam campaigns during November 2017 and June 2018. Single-photon time resolutions ranging from $104.3\,\rm{ps}$ to $114.8\,\rm{ps}$ and $83.8\,\rm{ps}$ to $112.7\,\rm{ps}$ have been achieved for MCP-PMTs with granularity $4 \times 64$ and $8 \times 64$ pixels, respectively. Photon yields are measured to be within $\sim$10\% and $\sim$30\% of simulation, respectively. Finally, the outlook for future work with planned improvements is  presented.

\end{abstract}

\maketitle

\section{Introduction}
\label{sec:Introduction}
\begin{figure}[ht]
\centering
\includegraphics[width=\linewidth]{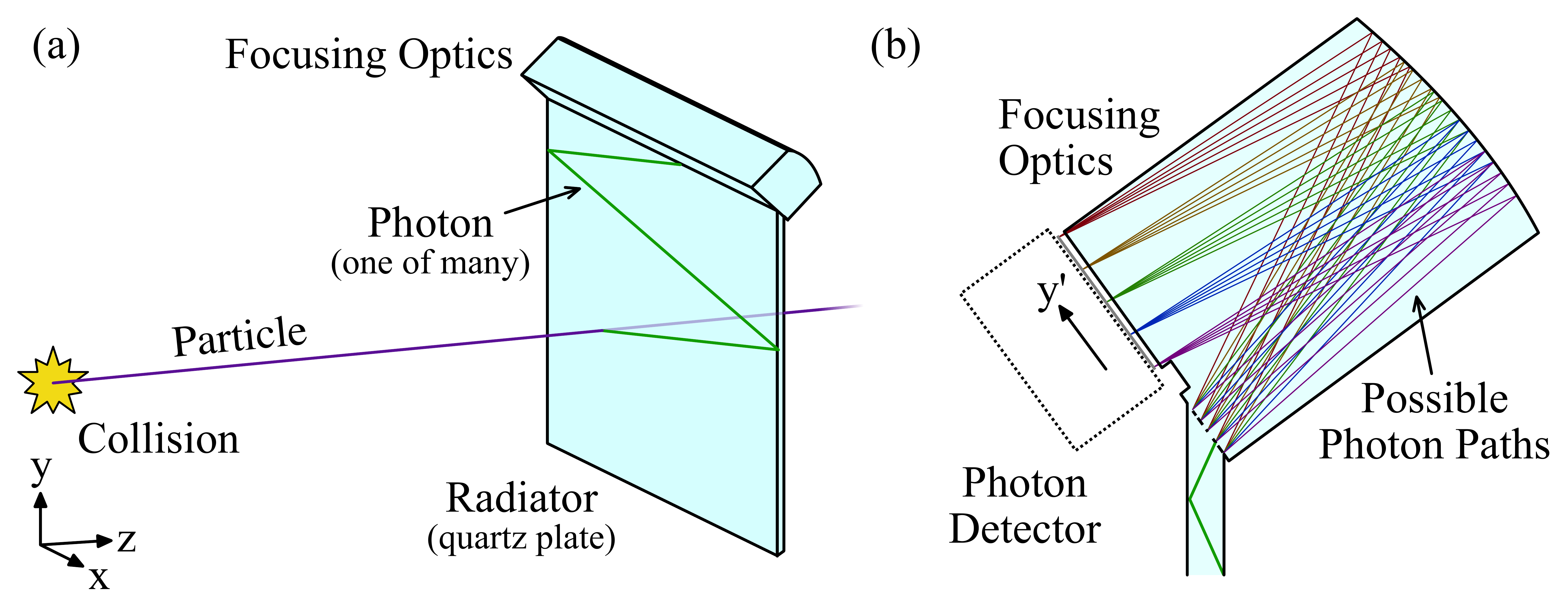}
\caption{
Schematics of a TORCH module demonstrating the principle of operation.
(a) Total internal reflection traps Cherenkov light generated by a particle traversing the radiator plate. (b) Upon reaching the focusing optics, the angle of the photon in the $y-z$ plane is mapped to the $y^{\prime}-$coordinate on the detector, allowing $\theta_C$ to be determined.
Note that the  $y^{\prime}$ axis is rotated by $36^{\circ}$ from the vertical ($y-$axis).
}
\label{fig:DetectorDesign}
\end{figure}

TORCH is a time-of-flight (ToF) detector designed to perform Particle IDentification (PID) at low momentum ($2-10\,\rm{GeV/\textit{c}}$) over a 10\,m flight path 
\cite{Charles_2011, Brook_2018}.  The principle of operation is demonstrated in \fig{fig:DetectorDesign}. TORCH exploits prompt Cherenkov photons produced by charged particles traversing a quartz plate of $10\,\rm{mm}$ thickness, combining timing measurements with DIRC-style reconstruction, a technique pioneered by the BaBar DIRC \cite{Adam_2005} and Belle II TOP \cite{Abe_2010, Fast_2017} collaborations.
A fraction of the radiated photons are trapped by total internal reflection, which then propagate to focusing optics at the periphery of the plate. Here a cylindrical mirrored surface maps the photon angle to a position on a photo-sensitive detector; custom-designed Micro-Channel Plate Photo-Multiplier Tube (MCP-PMT) detectors  \cite{Conneely_2015} are used to measure the times of arrival and positions of each photon. Combined with external tracking information, the spatial measurement allows the Cherenkov angle of the emitted photon to be determined.

The TORCH detector has been proposed for Upgrades Ib and II of the LHCb experiment in order to improve the pion, kaon, and proton separation capability of the experiment in the $2-10\,\rm{GeV/\textit{c}}$ range \cite{LHCb_Collab_2018}.
When installed in LHCb, TORCH will consist of eighteen identical $660 \times 2500 \times 10\,\rm{mm^3}$ modules located roughly $9.5\,\rm{m}$ from the interaction region. Over this distance the difference in time of flight between pions and kaons is $\sim35\,\rm{ps}$ for a momentum of $10\,\rm{GeV/\textit{c}}$, requiring a $10 - 15\,\rm{ps}$ time resolution for clean separation. This requires a single-photon timing resolution of $70\,\rm{ps}$, given around 30 detected photons per track.

This paper builds upon the work presented in Ref.~\cite{Brook_2018}, in which a small-scale TORCH demonstrator was tested with a  prototype  MCP-PMT of circular construction, produced by Photek, UK. The same demonstrator has now been verified with square 2-inch (nominal)  Photek tubes, in two test-beam campaigns at the CERN PS (East Hall T9 facility) during November 2017 and June 2018. The TORCH demonstrator is described in \secref{sec:TheTORCHDemonstrator}. The test-beam  infrastructure is presented in \secref{sec:TestbeamSetup}. \secref{sec:Calibrations} discusses the set of data-driven calibrations which were applied to the data. Results for the single-photon timing resolution and photon counting efficiency are presented in  \secrefs{sec:SinglePhotonTimeResolution}{sec:PhotonCounting}, respectively. Finally a summary and outlook for the future is given in \secref{sec:Summary}.

\section{The TORCH Demonstrator}
\label{sec:TheTORCHDemonstrator}

\subsection{Mechanics and Optics}
\label{subsec:MechanicsAndOptics}
The demonstrator consists of a $120$ (width) $\times 350$ (height) $\times 10$ (thickness) $\,\rm{mm^3}$ radiator plate, optically coupled to a focusing block which has a cylindrically mirrored surface designed to focus 2\,mm beyond the exit surface onto the MCP-PMT photocathode. The block has the same dimensions as it would have for a full-sized module in LHCb, except having its width reduced to match the 120\,mm width of the plate. The radiator plate and focusing block assembly was mounted into a rigid frame which allowed the angle of incidence of the beam to be varied
by tilting the demonstrator about the $x$-axis, seen in \fig{fig:DetectorDesign}. The complete structure was contained within a light-tight box and mounted upon a translation table, allowing the module to be positioned in the $x$ and $y$ directions with respect to the beam. Further details of the optical components and mounting mechanics can be found in Ref.~\cite{Brook_2018}.

\subsection{MCP-PMTs and Electronics}
\label{subsec:MCPsAndElectronics}
In each of the two test-beam campaigns, the demonstrator was instrumented with a different two-inch square MCP-PMT with a $64 \times 64$ anode pixelisation. The tubes were custom-designed for the TORCH project by Photek Ltd (UK) \cite{Conneely_2015} and represent the final prototypes of a three-stage development process \cite{Harnew_2018_2}. Charge from the MCP electron avalanche is collected on a resistive layer (``sea'') inside the PMT vacuum, and capacitively coupled to the anode pads. This allows charge sharing to improve the spatial resolution beyond the anode-pad pitch of 0.828\,mm.
In November 2017, the implemented MCP-PMT  had a $4 \times 64$ granularity in ($x, y^{\prime}$), where the coarse granularity was achieved by electrically grouping pixels on an external Printed Circuit Board (PCB), connected to the anode pads using anisotropic conductive film. In June 2018 the granularity in the $x-$direction doubled to $8 \times 64$, which, with charge sharing, gives an effective pixelisation which exceeds that required for optimal TORCH performance \cite{Garcia_2016}. For LHCb installation, a pixelisation of $8 \times 128$ is planned.

Both MCP-PMTs have an active area of $53 \times 53 \,\rm{mm^{2}}$, corresponding to approximately half the width of the demonstrator. In both test-beam campaigns the MCP-PMT was mounted between one side edge and the centre of the focusing block, with the other half of the detector plane not being instrumented.

\begin{figure}[h]
\centering
\includegraphics[width=0.75\linewidth]{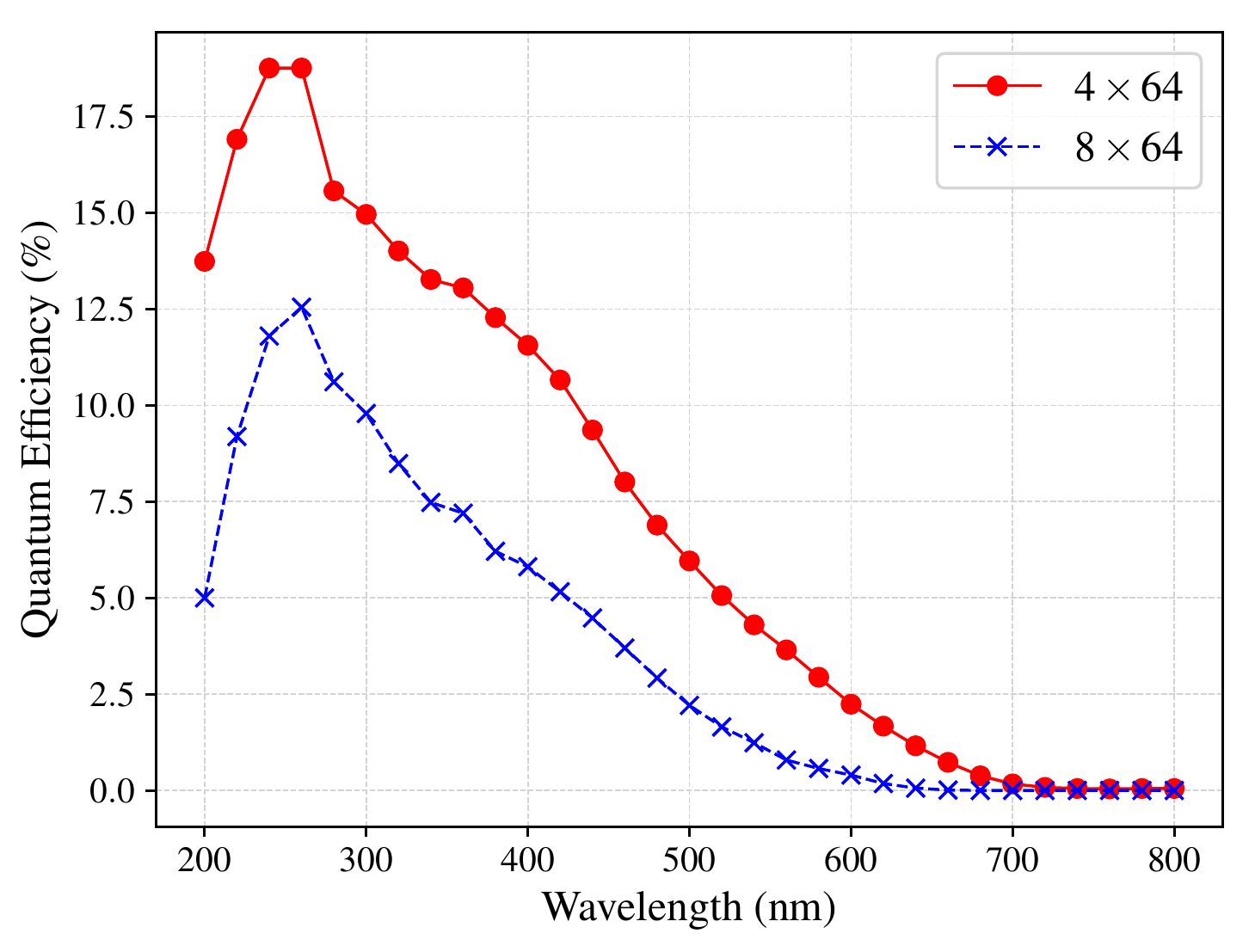}
\caption{The quantum efficiencies of the $4 \times 64$ (November 2017) and the $8 \times 64$ (June 2018) MCP-PMTs, measured at CERN.}
\label{fig:QE}
\end{figure}

In the $4 \times 64$ MCP-PMT, the insulating layer which separates the resistive-sea from the anode readout pads has a thickness of 0.5\,mm. This results in a point-spread function at the pads of 1.80 $\pm$ 0.15\,mm (FWHM), which was determined from laboratory measurement and verified by simulation \cite{Conneely_2015}.
The $8 \times 64$ MCP-PMT has a 0.3\,mm insulating layer, and results in a point-spread function at the anode pads of 1.30 $\pm$ 0.13\,mm (FWHM).
The quantum efficiencies (QEs) of both tubes were measured in the laboratory, and are shown in \fig{fig:QE}. It can be seen that the integrated QE of the $8 \times 64$ MCP-PMT is around a factor two less than for the $4 \times 64$ device. Although the QE of the $8 \times 64$ is not optimal for reaching the desired number of photons per track, the performance of future tubes is expected to improve with further iterations of development.

Readout electronics employing the NINO \cite{Anghinolfi_2004} and HPTDC \cite{Akindinov_2004} chipsets were custom-developed for the TORCH project \cite{Gao_2016}. Due to the increased granularity of the $8 \times 64$ MCP-PMT in the coarse-pixel direction, an entirely new readout system was developed  to replace that used for the $4 \times 64$ device. Because of differences in the size and shape of the boards, new holding mechanics were fabricated for the  $8 \times 64$ device, which introduced a $5\,\rm{mm}$ upwards offset of the MCP-PMT in the $y^{\prime}-$direction relative to the  $4 \times 64$ device.

\subsection{Hit Clustering}
\label{subsec:HitClustering}
As previosuly discussed, the Photek MCP-PMT was designed so that a single incident photon will give hits on several neighbouring pixels \cite{Conneely_2015}. This means that the $64$ physical pixels in the $y^{\prime}-$ direction can provide an effective granularity of $128$ pixels by exploiting charge sharing. In this way, to reconstruct single photons, hits are clustered according to the following criteria:
\begin{itemize}
\item they must have the same $x-$coordinate (coarse pixel direction);
\item they must be adjacent neighbours in the $y^{\prime}-$coordinate;
\item the arrival of the hit must be timed within $1\,\rm{ns}$ of its neighbour.
\end{itemize}
All three criteria must be met for any pair of hits to be included in the same cluster.
However for the $8 \times 64$ dataset, the criteria were slightly modified to account for a small fraction of dead channels: namely, if two clusters fall on either side of a known dead channel and the hits neighbouring the dead channel fall within $2\,\rm{ns}$ of each other, then the clusters are merged. Cluster size is determined by the number of hits in the cluster, and the cluster position is taken to be the average position of the centroid of all the hits.

\subsection{Detector Simulation}
\label{subsec:Simulation}
A simulation of the TORCH demonstrator has been developed, using optical processes modeled by Geant4 \cite{Allison_2016}. Custom libraries were used to model the detector response and readout, which take input from laboratory measurements including the MCP-PMT quantum efficiency, gain, and point-spread function.
Losses due to quartz surface scattering and Rayleigh scattering are modelled.
The same simulation was used for both test-beam periods, but with differing input from laboratory-measured parameters  for the respective MCP-PMT used.

\section{Test-beam Setup}
\label{sec:TestbeamSetup}
In both test-beam campaigns, a $5\,\rm{GeV/\textit{c}}$ beam was used, comprising approximately 70\% pions and 30\% protons.
The TORCH demonstrator was positioned with the beam striking half way down the radiator plate, $5\,\rm{mm}$ from the edge (below the MCP-PMT), and tilted back from the vertical by $5^{\circ}$. This geometrical configuration  ensured that the Cherenkov pattern was well contained on the MCP-PMT detector surface.

\begin{figure}[h]
\includegraphics[width=0.95\linewidth]{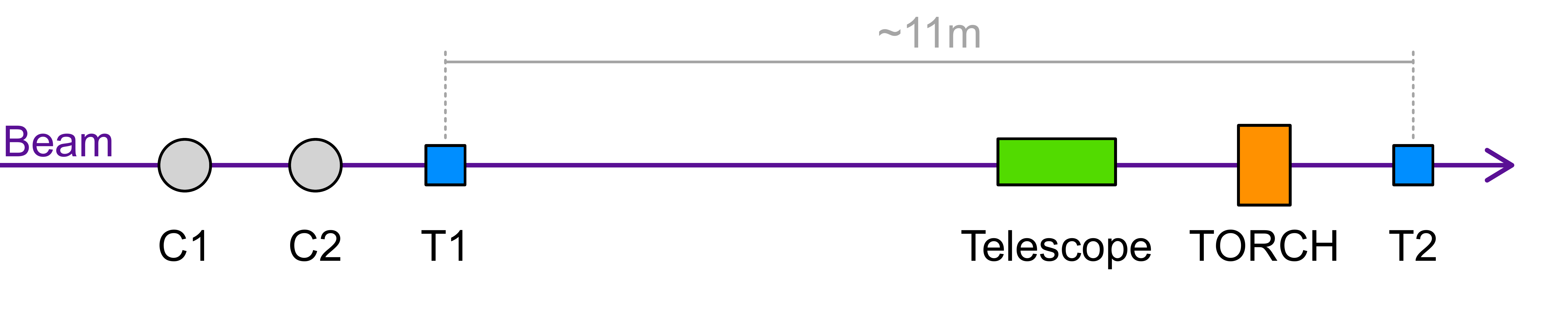}
\caption{A schematic showing the beam-line configuration. C1 and C2 are Cherenkov counters. T1 and T2 are timing stations spaced approximately $11\,\rm{m}$ apart.}
\label{fig:TestbeamSetup}
\end{figure}

\begin{figure}[h]
\centering
\includegraphics[width=0.75\linewidth]{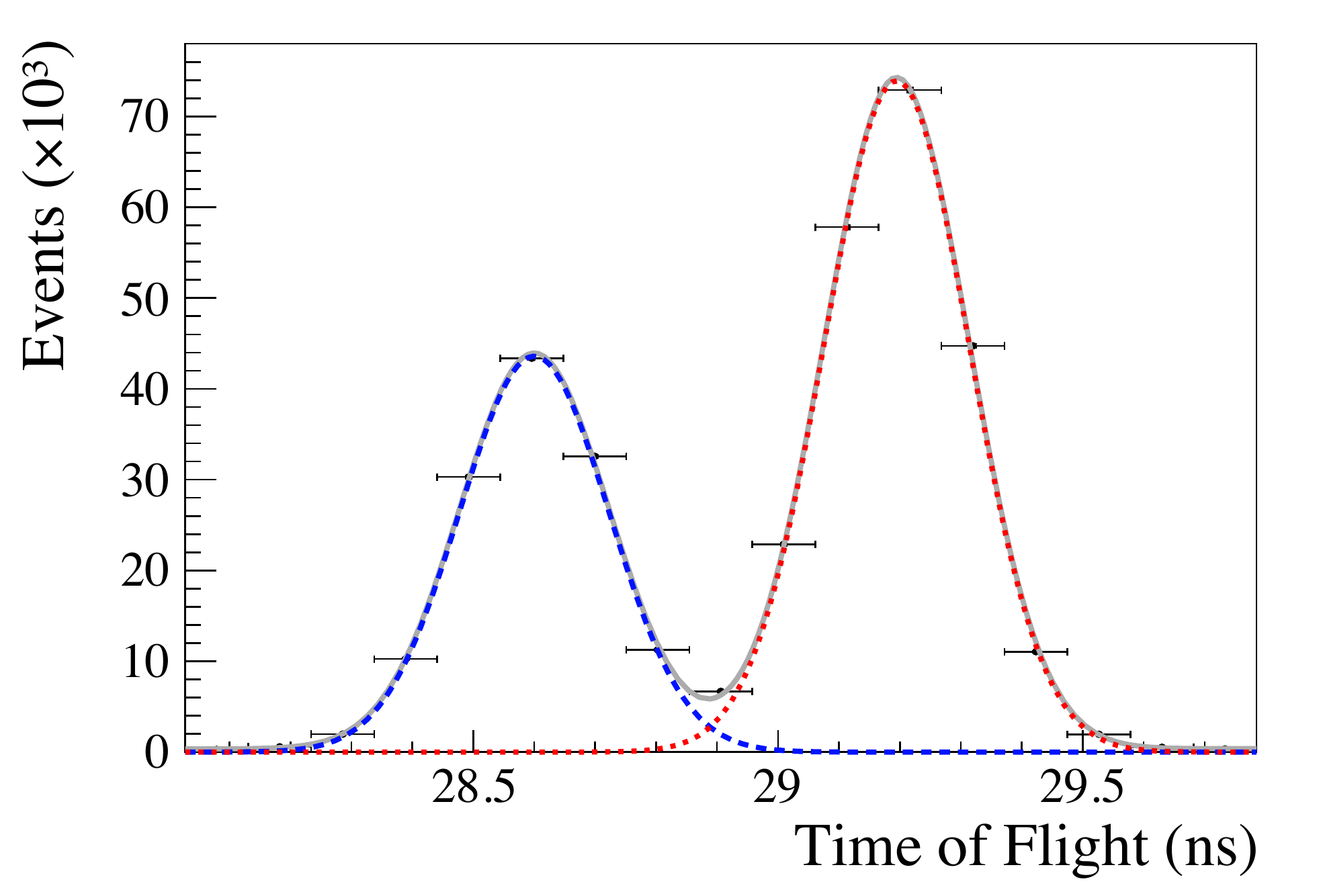}
\caption{The time-of-flight difference (T2$-$T1) over an 11\,m flight path, showing the pion contribution in red (dotted) and the proton contribution in blue (dashed). Note that the zero time is arbitrary and, as defined, pions arrive later in time than protons. The horizontal bars denote the bin width.}
\label{fig:TimeRef}
\end{figure}

The same beam-line  infrastructure was installed for both campaigns, displayed schematically in \fig{fig:TestbeamSetup}.
A pair of identical timing stations, T1 and T2, spaced approximately $11\,\rm{m}$ apart, was used to provide a time reference for TORCH. Each station, oriented at 49$^\circ$ to the beam, consisted of a 100\,mm long,
8$\times$8\,mm$^2$  borosilicate bar in which Cherenkov light was generated from traversing particles.
A single-channel MCP-PMT detected the direct photons and provided a precise timing signal. The signals were injected into the TORCH electronics and read out simultaneously with the rest of the data. By combining signals from both stations, a time of flight measurement could be made independently of TORCH, providing a cross-check of PID for the particle traversing the TORCH prototype. Additionally, each station had a pair of scintillators providing an $8 \times 8\,\rm{mm^2}$ coincidence. Requiring a signal in both scintillators narrowed the beam definition accepted by the trigger and improved the resolution of the time reference.  The timing power  of the stations is demonstrated in \fig{fig:TimeRef}, which shows clearly the separation of pions and protons in the beam.
In addition, a pair of threshold Cherenkov counters filled with $\rm{CO_{2}}$ at $2.5\,\rm{bar}$ were introduced for both campaigns, and provided the independent source of PID.

\begin{figure}[h]
\centering
\includegraphics[width=0.75\linewidth]{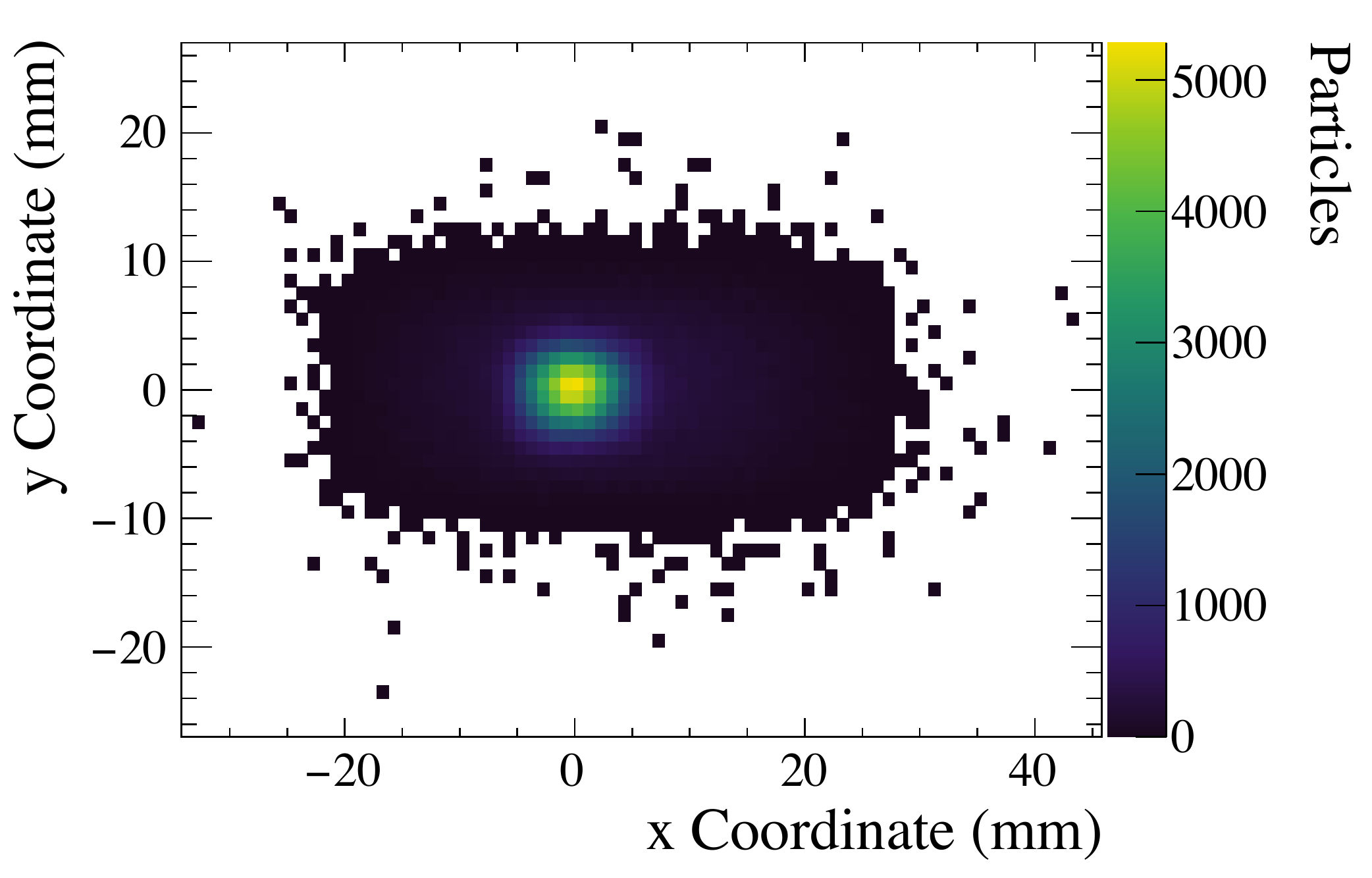}
\caption{The beam profile in $x$ and $y$ measured by the telescope when extrapolated to the TORCH radiator. The outer structure of the shape maps the beam profile and is attributed to scattering. The central distribution is the region defined by the scintillator trigger.}
\label{fig:Beamprofile}
\end{figure}

An EUDET/AIDA pixel beam telescope \cite{Rubinskiy_2012} was also installed in the beam-line, consisting of six $18.4\,\rm{\mu m}$ pitch sensors (Mimosa26).
The telescope allowed an accurate measurement of the beam profile incident on TORCH, even though an event-by-event synchronization was not possible.
\fig{fig:Beamprofile} shows the beam profile measured by the telescope when extrapolated to the TORCH radiator, giving an RMS spot size of $2.73 \pm 0.02$\,mm in $x$ and $2.01 \pm 0.02$\,mm in $y$.  The beam divergence was measured to be $5.8 \pm 0.2$\,mrad and $2.6 \pm 0.2$\,mrad in $x$ and $y$, respectively.

Triggering of the TORCH readout and telescope was provided by an AIDA-2020 Trigger Logic Unit (TLU) \cite{Baesso_2019}.
The new beam-line infrastructure allowed a large increase in achievable data rate with respect to \cite{Brook_2018}.
By providing the independent source of PID, the Cherenkov counters allowed T1 to be removed from the trigger, with T2 alone being used as a time reference. This led to a wider beam profile that could be triggered upon, significantly increasing the acceptance and trigger rate.
Comparing the PID information from the Cherenkov counters with the PID from ToF, the purities of the pion and proton samples from the Cherenkov counters were approximately 94\% and 82\% in November 2017, and 98\% and 96\% in June 2018, respectively.

\section{Calibrations}
\label{sec:Calibrations}
Two data-driven calibrations were applied to the data to correct the timing of the MCP-PMT output signals, the first to account for timewalk in the NINO chip, and the second to correct for integral non-linearity in the HPTDC.


The first correction accounts for  timewalk of the NINO (i.e.~differences in timing due to variations of pulse amplitude), and is adapted from the data-driven method employed in Ref.~\cite{Brook_2018}.
The first stage in the calibration process is to define an MCP-PMT photon cluster.
Assuming each cluster corresponds to a single photon, the hit pixels within that cluster should have simultaneous recorded times, and any difference $\Delta t_{i,j}$ between pairs of channels $i, j$ would be a consequence of time slew. The NINO utilizes a time-over-threshold technique, outputting a binary signal with a width defined by the rising and falling edges of the MCP-PMT  input pulse when passing an adjustable threshold. The signal width $w_i$ for a channel $i$
is hence related to the input pulse amplitude. This introduces a relationship between $\Delta t_{i,j}$ and the corresponding pulse widths, which can be parameterised as:
\begin{equation}
\Delta t_{i,j} = t_{i} - t_{j} = F \left( w_{i} , w_{j}\right),
\end{equation}
where $t_{i,j}$ are the recorded times of the hits on channels $i,j$, and $F$ is chosen to be a 2-dimensional function of quadratic form, with those coefficients determined by a fit to pairs of hits from the same cluster.
The constant term also corrects for the relative delay between individual pixel timing offsets ($t_{0}$'s).
In this way, the parameters of $F$ can be determined for all pairs of channels during a data run. Thereafter a correction is made to the measured arrival time of each single pixel hit according to its measured pulse width.
This method assumes the time walk of each individual pixel is uncorrelated with all the others, and improves the method employed in Ref.~\cite{Brook_2018} by comparing all pairs of hits, rather than parameterising and correcting only next-to-nearest neighbours.

The second calibration accounts for non-linearity in the HPTDC chip, where the bins used to digitise the data are not equally spaced in time \cite{Akindinov_2004}, leading to integral non-linearity. Several large dedicated calibration datasets were taken to allow a code-density test \cite{Liu_2010} to be performed to correct for this effect.

A calibration step which is presently missing is the so-called charge-to-width calibration, which would allow a more accurate measurement of the amount of charge collected in any given pixel hit
as a function of the width of the output pulse. This in turn would allow more accurate cluster centroiding to be performed. This calibration requires a dedicated  laboratory-based  charge-injection system, and this is currently under development.

\section{Single-Photon Time Resolution}
\label{sec:SinglePhotonTimeResolution}

\begin{figure}[htb]
\centering
\includegraphics[width=0.75\linewidth]{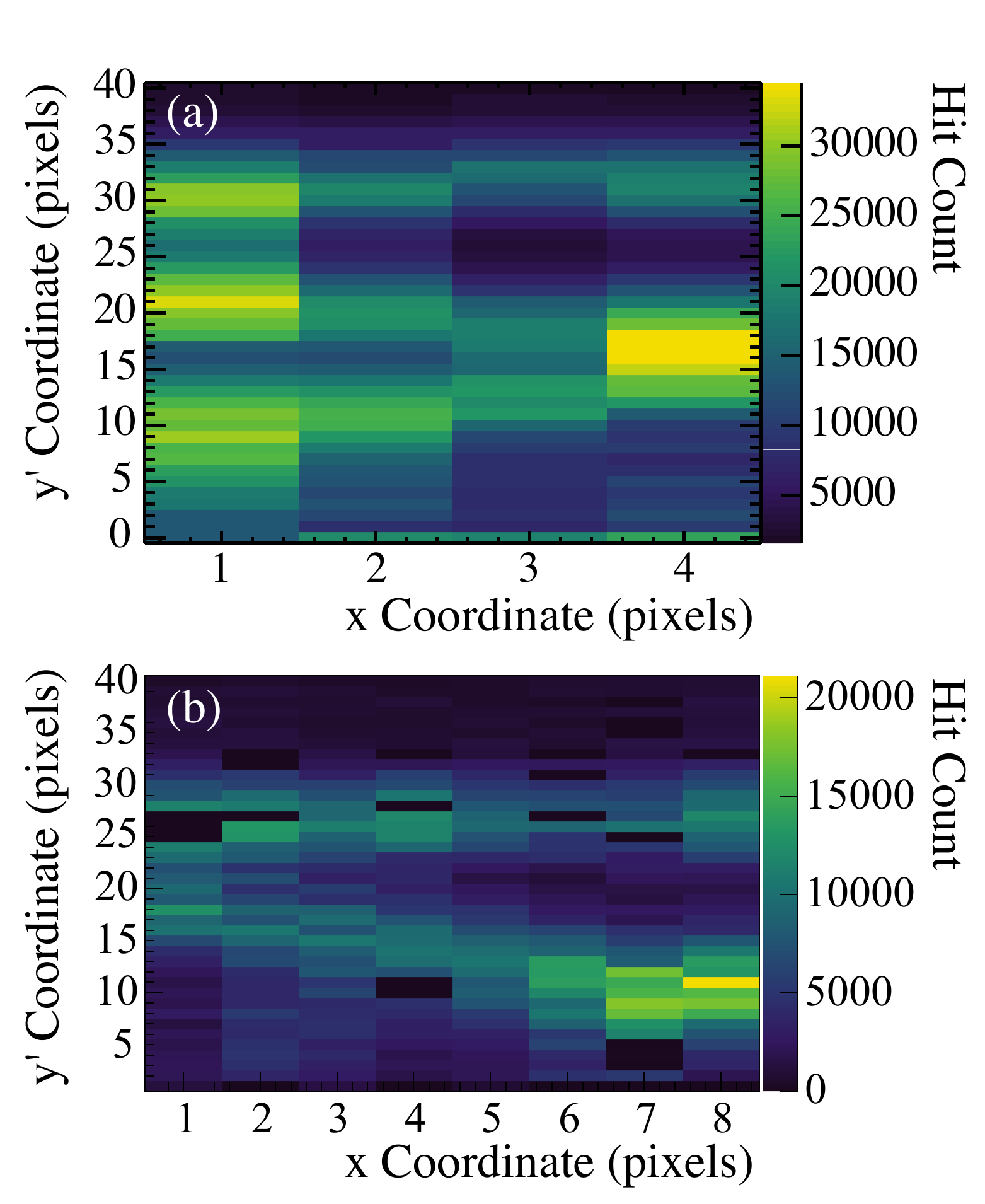}
\caption{Uncorrected MCP-PMT hit-maps for combined pions and protons showing (a) the $4 \times 64$  and (b) the  $8 \times 64$ MCP-PMT (November 2017 and June 2018 test-beam periods, respectively). The black bins in (b) indicate a dead channel. Note that although both beam tests used an MCP-PMT with 64 pixels in the $y^{\prime}$ coordinate, the top part of the detector was not illuminated and so the distributions have been truncated.}
\label{fig:Hitmaps}
\end{figure}

Figure \ref{fig:Hitmaps} shows the uncorrected distributions (pixel maps) of hits on the MCP-PMTs
from the two run periods for pions and protons combined, taken with the beam positioned at the vertical mid-point of the radiator plate, $5\,\rm{mm}$ from the edge below the MCP-PMT. Bands can be seen, corresponding to different photon paths within the radiator plate.
The empty bins in \fig{fig:Hitmaps}b indicate dead channels. These are attributed to broken wire bonds of the NINO electronics board, an issue which has been resolved in subsequent iterations.

The single-photon time resolution of the demonstrator can be measured by comparing the time at which a photon is detected to that predicted from the TORCH  reconstruction algorithm \cite{Brook_2018}. The algorithm determines  the photon path  in the radiator plate and the Cherenkov angle from the position of the track entry point, the track direction, and the position of the photon hit on the MCP-PMT. Combining this with knowledge of the primary particle species and its momentum, the time of propagation can then be calculated through the intermediate steps of determining the phase and group refractive indices. Note that the nominal values of beam position and incident angle are used in the reconstruction, leaving any finite beam width and divergence to be accounted for statistically, as described below.

\begin{figure}[h]
\centering
\includegraphics[width=0.75\linewidth]{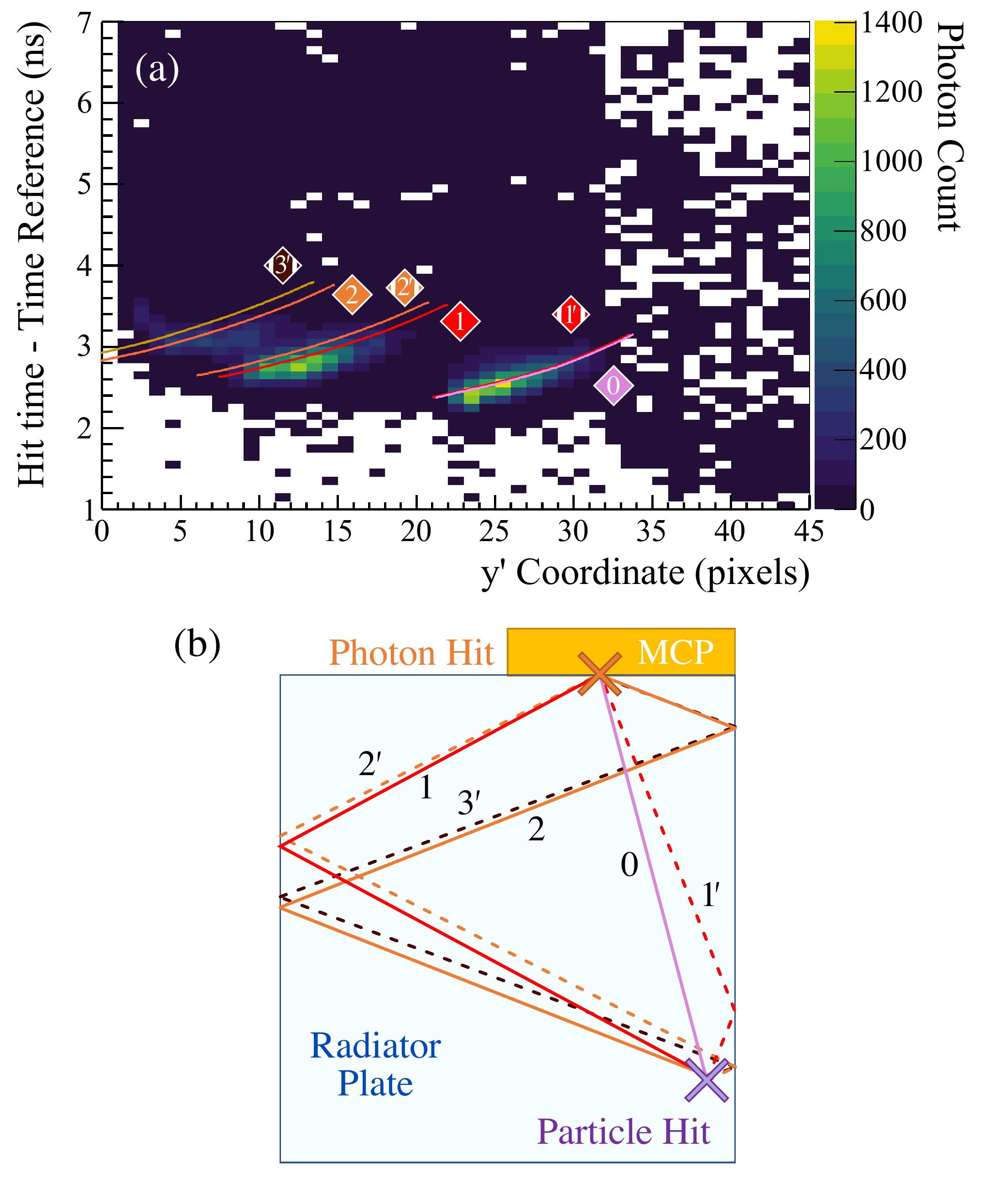}
\caption{
(a) The distribution of detected photons as a function of $y^{\prime}$ (finely-granulated) pixel number and arrival time, for column 5 of the $8 \times 64$ MCP-PMT (see \fig{fig:Hitmaps}b). Here the selection is for protons.  The three distinct peaks correspond to different orders of reflection off the sides of the radiator plate, and the overlaid lines show the predicted time of arrival  for each order. The reduced population of hits for pixels beyond 32 is due to the different threshold settings for a pair of adjacent NINO chips which read out the MCP-PMT column.
(b) A schematic of the photon paths assigned in the reconstruction which correspond to the overlaid lines in (a), labelled according to the number of photon side reflections. Paths first reflecting off the edge closest to the beam are shown by dotted lines, and their labels have primes.
}
\label{fig:TimeProjectionExample}
\end{figure}

For each column of pixels, the measured arrival time can be plotted against the $y^{\prime}$ (finely-granulated) pixel number. \fig{fig:TimeProjectionExample}a shows an example distribution for the $8 \times 64$ MCP-PMT, selected for protons only, with the predictions from the reconstruction algorithm overlayed. In calculating the predicted time in the reconstruction, each photon is treated individually and its energy calculated \cite{Charles_2011}.
The distinct bands seen in the figure correspond to the different orders of reflection from the side faces of the demonstrator, illustrated schematically in  \fig{fig:TimeProjectionExample}b.
This clearly demonstrates that photon paths in the radiator plate are well separated. Note that within a given order of reflection for a specific set of track parameters, the measured $y^{\prime}$  pixel coordinate is correlated to the Cherenkov photon energy, with the finite pixel size contributing to the chromatic uncertainty.

\FloatBarrier

For those photon hits which had either no reflections off a side edge or which only had a reflection off the edge below the MCP-PMT (corresponding to orders $0$ and $1^{\prime}$ in \fig{fig:TimeProjectionExample}b), a residual distribution (i.e.~the measured minus predicted times of arrival) is constructed. The residual distributions of individual bins in $y^{\prime}$ are first fitted to determine the resulting mean. These means are expected to be offset with respect to each other due to chromatic dispersion, and thus are corrected by offsetting the photon arrival times within each bin by the mean of the bin. Recombining the bins then gives the final fitted distribution. The sigmas are also dependent on photon energy, hence the measurements are averaged.

\FloatBarrier

\begin{figure}[h]
\centering
\includegraphics[width=0.75\linewidth]{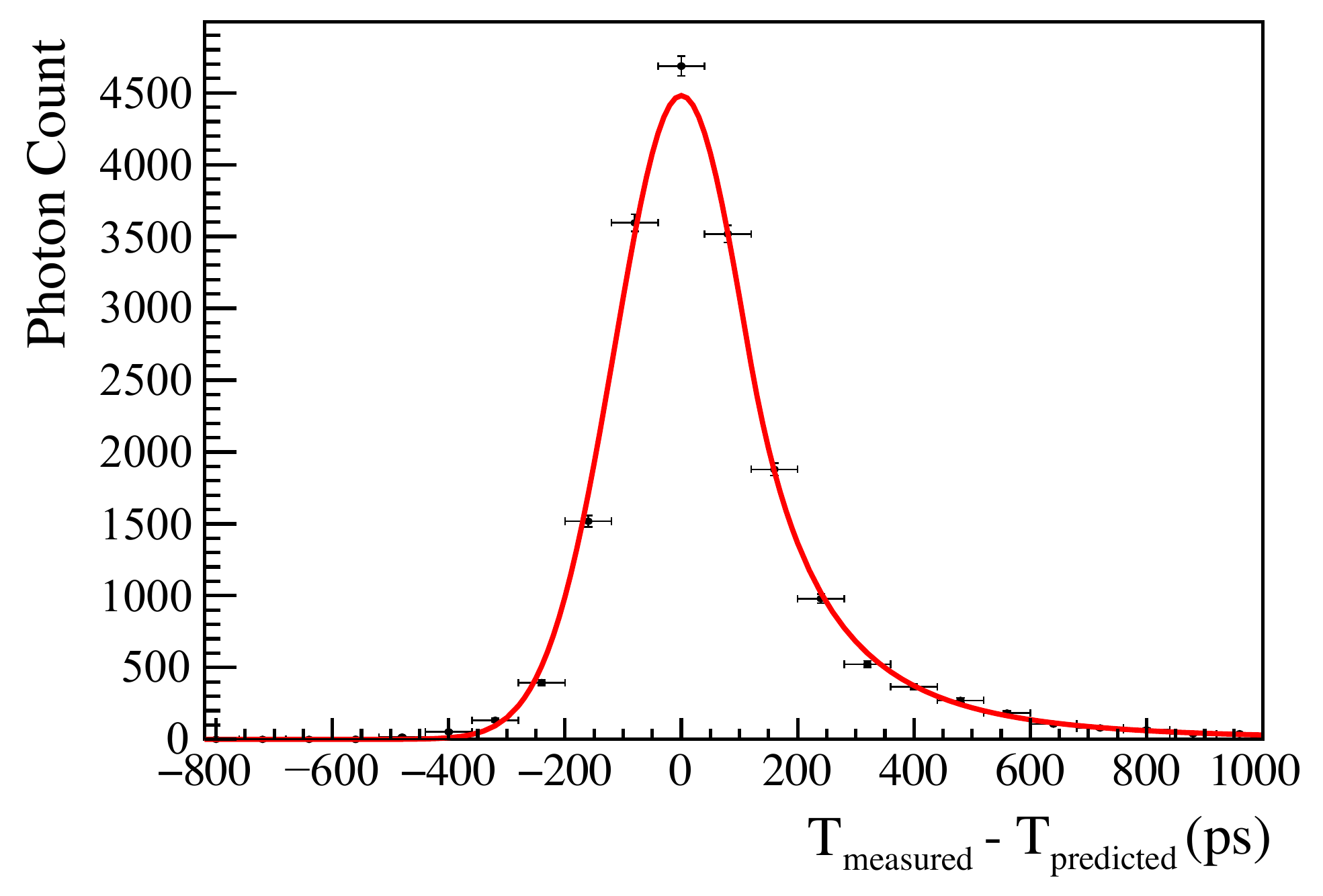}
\caption{An example residual distribution for orders of reflection $0$ and $1^{\prime}$, as indicated in \fig{fig:TimeProjectionExample}b. The data are fitted with a Crystal Ball function with parameters $\mu = 0 \pm 1\,\rm{ps}$, $\sigma = 115 \pm 1\,\rm{ps}$, $\alpha = -1.01 \pm 0.03$, and $n = 4.5 \pm 0.4$ \cite{Skwarnicki_1986}. The contribution from the timing reference has not yet been subtracted.
The tail is attributed to backscattering of primary photoelectrons in the MCP-PMT and possibly a small contribution from residual timewalk.
}
\label{fig:TimingResidualExample}
\end{figure}

Figure \ref{fig:TimingResidualExample} shows an example of a residual distribution fitted with a ``Crystal Ball'' function, consisting of a Gaussian core with a power-law tail \cite{Skwarnicki_1986}. The tail models back-scattering of primary photoelectrons in the MCP-PMT and possibly a small contribution from residual timewalk, whilst the standard deviation of the Gaussian component is interpreted as the measured time resolution.

The time spread of the residual distribution, $\sigma_{\rm{Total}}$, is a combination of several factors which must be subtracted in quadrature to give the intrinsic timing resolution. This is given by
\begin{equation}
\label{eqn:TimeSpread}
\sigma_{\rm{Total}}^{2} = \sigma_{\rm{TORCH}}^{2} + \sigma_{\rm{Beam}}^{2} + \sigma_{\rm{TimeRef}}^{2},
\end{equation}
where $\sigma_{\rm{TORCH}}$ is the intrinsic single-photon timing resolution which we wish to determine, $\sigma_{\rm{Beam}}$ is the spread in the residual distribution resulting from the finite width
and divergence of the incident beam, and $\sigma_{\rm{TimeRef}}$ is the time resolution of the T2 station that provides the reference time for TORCH. The latter two contributions are discussed below.

{\it The beam-spread contribution}:
The contribution to the time spread due to the beam profile, $\sigma_{\rm{Beam}}$,
is determined from simulation.
The simulation is run with the spread in beam position and divergence as measured by the telescope, and then for a beam with no spread.
 In each case, a residual distribution is constructed the same way as for data, and the width of the two compared.
The value of $\sigma_{\rm{Beam}}$ is found to be $14\pm 2\,\rm{ps}$, where the quoted error accounts for the uncertainty on the beam profile and differences between the values measured for each MCP-PMT column.


{\it The time reference contribution}:
The  T2  station provides the time reference with respect to which the Cherenkov photons in TORCH are measured.
The   time resolution of the stations is demonstrated in \fig{fig:TimeRef}, showing the (T2$-$T1) time difference in June 2018 data. The pion and proton beam contributions are clearly separated.
To determine the resolution of  the downstream station T2 independent of TORCH, it is assumed the behaviour of the pair of time reference stations T1 and T2 is identical before their signals propagate to the TORCH readout, however the timing of  upstream station T1 is degraded due to $\sim$11 m of cable length.
When this contribution is isolated, the resolution of T2 is determined to be  $\sigma_{\rm{TimeRef}} =   43.4 \pm 1.5$\,ps (pions)  and $42.8 \pm 2.0$\,ps (protons).


Using Equation \eqref{eqn:TimeSpread}, the time resolution $\sigma_{\rm{TORCH}}$ is determined separately for each MCP-PMT column and for each incident particle species. Unfortunately a significant pollution of pions is observed in the proton sample for the November 2017 dataset due to the Cherenkov counters being non-optimally tuned, so only photons resulting from an identified pion are used in this case.
This results in four measurements for the $4 \times 64$ MCP-PMT, presented in \tab{tab:Nov17Results}, and 16 for the $8 \times 64$, shown in \tab{tab:Jun18Results}.

\begin{table}[htb]
\centering
\begin{tabular}{c|c}
MCP Column & $\sigma_{\rm{TORCH}}\,\rm{(ps)}$ \\
\hline
   & \\
1 & \errval{112.1}{1.4} \\
2 & \errval{114.8}{1.4} \\
3 & \errval{104.3}{1.4} \\
4 & \errval{111.4}{1.4}
\end{tabular}
\caption{The single-photon time resolutions for pions measured for the $4 \times 64$ MCP-PMT in November 2017. The MCP column numbers match those shown in \fig{fig:Hitmaps}a. The quoted uncertainties are purely statistical.}
\label{tab:Nov17Results}
\end{table}

\begin{table}[htb]
\centering
\begin{tabular}{c|c|c}
MCP & $\sigma_{\rm{TORCH}}$ & $\sigma_{\rm{TORCH}}$ \\
Column & Pions$\,\rm{(ps)}$ & Protons$\,\rm{(ps)}$ \\
\hline
 &  & \\
1 & \errval{110.6}{1.2} & \errval{112.7}{1.4} \\
2 & \errval{101.7}{1.2} & \errval{110.6}{1.4} \\
3 & \errval{101.5}{1.2} & \errval{110.6}{1.4} \\
4 & \errval{105.5}{1.2} & \errval{106.2}{1.4} \\
5 & \errval{83.8}{1.3} & \errval{91.0}{1.4} \\
6 & \errval{101.3}{1.2} & \errval{103.4}{1.2} \\
7 & \errval{90.3}{1.2} & \errval{87.5}{1.4} \\
8 & \errval{112.4}{1.1} & \errval{102.8}{1.4}
\end{tabular}
\caption{The single-photon time resolutions for pions and protons measured for the $8 \times 64$ MCP-PMT in June 2018. The MCP column numbers match those shown in \fig{fig:Hitmaps}b. The quoted uncertainties are purely statistical.}
\label{tab:Jun18Results}
\end{table}

The measurements from the two test-beam periods are generally similar, with resolutions $\sigma_{\rm{TORCH}}$ between 100 and 110\,ps typically observed. The overall trend of enhanced performance in the $8 \times 64$ MCP-PMT with respect to the $4 \times 64$ is attributed to the improved resolution in the $x-$direction due to the doubling of the number of pixel columns.
Columns 5 and 7 stand out in particular for the $8 \times 64$  MCP-PMT data, with measured  resolutions of order 90\,ps. This results from the  application of better calibration corrections for these columns. It is noted that six of the eight columns for the $8 \times 64$ MCP-PMT give better resolutions for pions than protons, an effect which is attributed to  a residual pion pollution in the proton sample  in the June 2018 data. In this case a fraction of pions will be falsely reconstructed as protons, resulting in an incorrect predicted time.

As indicated in \fig{fig:TimeProjectionExample}a, orders 0 and 1 cannot be distinguished in data. The difference in time of arrival between the two orders from the simulation ranges from $5 - 30\,\rm{ps}$, varying with the $x-$position of the hit. This will widen the residual distribution, and leads to a slightly degraded time resolution being measured than with only a single order of reflection. However, as the effect is photon-energy dependent, no attempt has been made to subtract this contribution.

Incorporating a charge-to-width calibration for the NINO in addition to the data-driven approach here employed would allow a charge-weighted average of the time and position of each cluster to be determined. This would improve the resolution further, and closer to the desired $70\,\rm{ps}$.

\FloatBarrier
\section{Photon Counting}
\label{sec:PhotonCounting}
\begin{figure}[htb]
\centering

\includegraphics[width=0.73\linewidth]{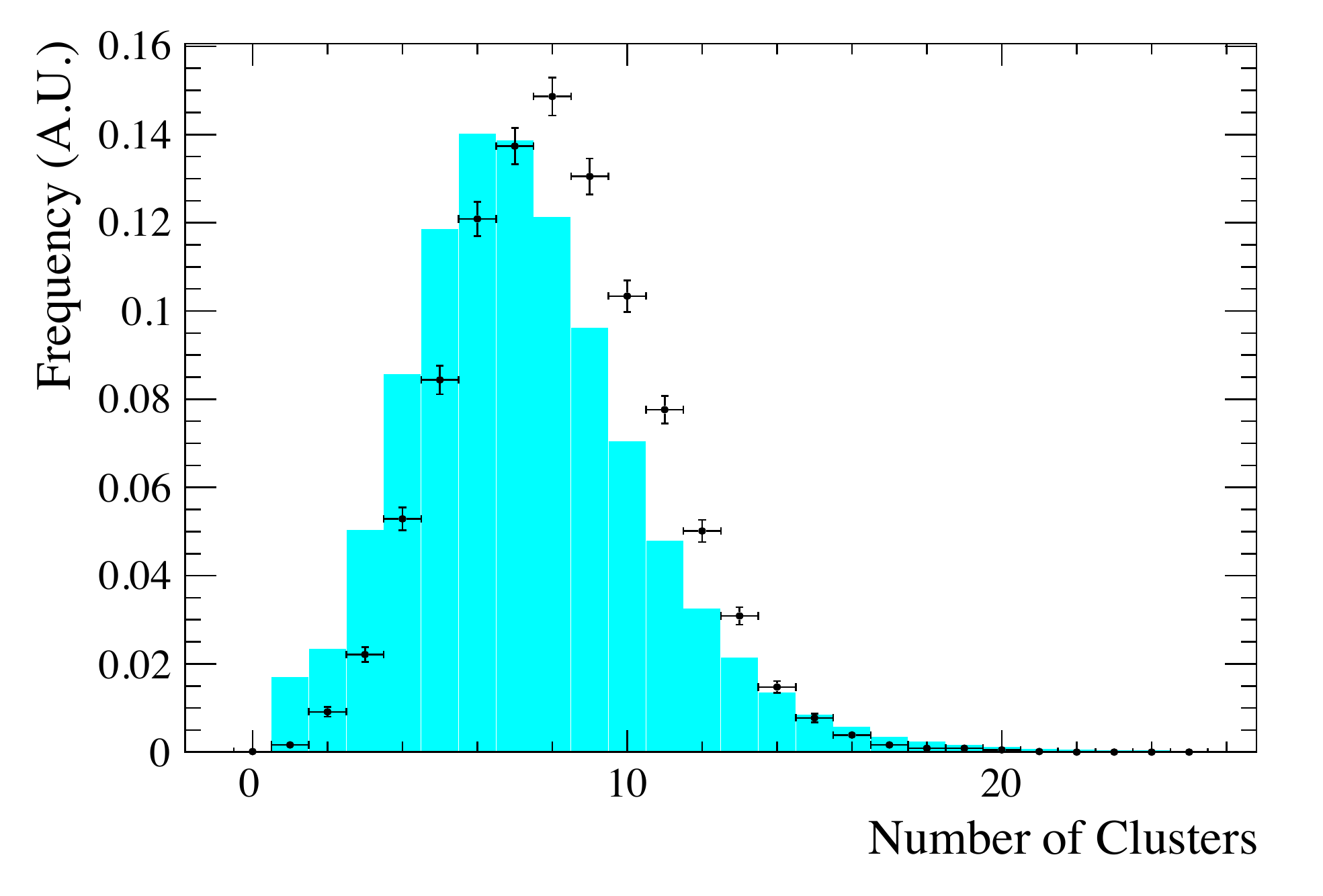}\put(-50,140){(a)}

\includegraphics[width=0.73\linewidth]{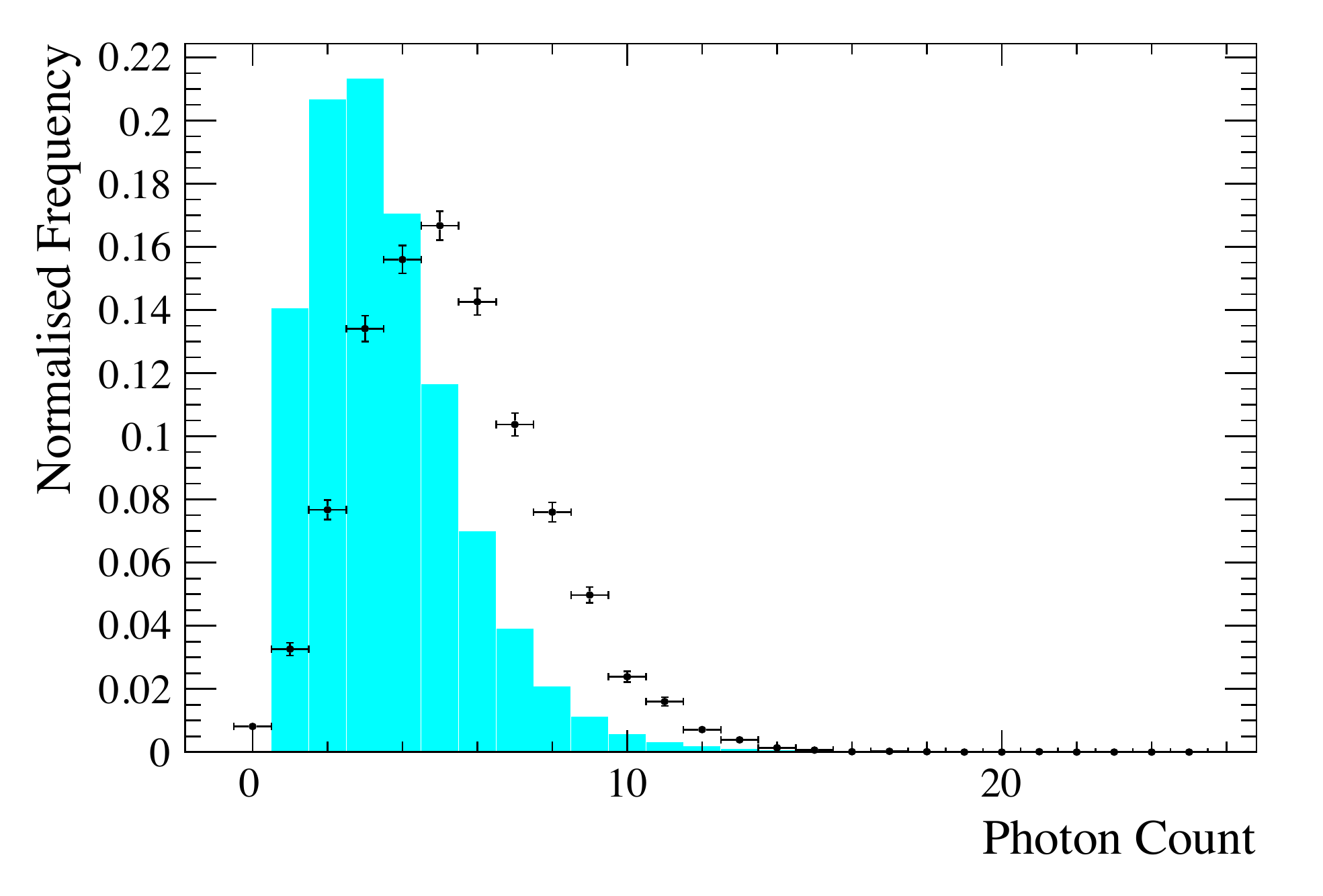}\put(-50,140){(b)}
\caption{The number of clusters seen in data (histograms) and simulation (points) for (a) the $4 \times 64$  and (b) the $8 \times 64$ MCP-PMT.
The smaller number of photons observed in simulation for  the $8 \times 64$  is due to its lower quantum efficiency.}
\label{fig:PhotonCounting}
\end{figure}

The photon counting efficiency of the demonstrator is determined by counting the number of detected clusters and  comparing to the number expected in simulation.
Monte Carlo samples corresponding to 10000 incident pions at 5\,GeV/\textit{c}  were used for each detector configuration.
Figure \ref{fig:PhotonCounting} shows the distributions of number of photons seen and expected in data and simulation, respectively, for the two MCP-PMT arrangements. The arithmetic mean numbers of measured  photons are compared in \tab{tab:PhotonCountingMeans}. A negligible difference is observed in counting efficiency between pions and protons, hence no selection is made in the data based on the species.

\begin{table}[htb]
\centering
\begin{tabular}{c|c|c|c}
MCP-PMT & Data & Simulation & Ratio \\
\hline
 &  &  & \\
$4 \times 64$& \errval{7.37}{0.03} & \errval{8.12}{0.03} & \errval{0.908}{0.005} \\
$8 \times 64$ & \errval{3.53}{0.02} & \errval{5.27}{0.03} & \errval{0.670}{0.005}
\end{tabular}
\caption{The arithmetic means of the photon-yield distributions shown in \fig{fig:PhotonCounting} and the ratio of data compared to simulation. The quoted uncertainties are purely statistical.}
\label{tab:PhotonCountingMeans}
\end{table}

The reduced number of clusters observed for the $8 \times 64$ MCP-PMT with respect to the $4 \times 64$ device in both data and simulation is expected, given that
the $8 \times 64$ device has a significantly lower quantum efficiency (as seen in \fig{fig:QE}).
The photon yield in the $4 \times 64$ MCP-PMT agrees within 10\% of the simulation, however for the $8 \times 64$ device, a $\sim$30\% loss in data is observed. The yields depend strongly on the MCP-PMT gains and the NINO thresholds, the best estimates of which were used in the simulation\footnote{The gain of the $4 \times 64$ has been measured to be $1.8 \times 10^{6}$ electrons, whilst the nominal value of the gain for the $8 \times 64$ is $1.0 \times 10^{6}$ electrons, taken from the Photek data sheet. These values were used respectively, along with a NINO threshold value of 30\,fC.}.
The observed discrepancies are attributed to uncertainties due to small signals arising from charge sharing, for which the systematics from the NINO threshold values are significant.
Future laboratory work will therefore focus on improving the efficiency and calibration of the MCP-PMTs and the electronics.


\FloatBarrier
\section{Summary and Future Plans}
\label{sec:Summary}
Studies of a small-scale TORCH demonstrator with customised MCP-PMTs and readout electronics  have been performed during two test-beam periods in  November 2017 and June 2018. Single-photon time resolutions ranging from $104.3\,\rm{ps}$ to $114.8\,\rm{ps}$ and $83.8\,\rm{ps}$ to $112.7\,\rm{ps}$ have been measured for MCP-PMTs of  granularity $4 \times 64$ and $8 \times 64$, respectively. The improvement for the $8 \times 64$ is attributed to its factor two increase in granularity. The measurements are within $30 - 40\%$ of the $70\,\rm{ps}$ targeted. The photon yields show a strong dependence on the MCP-PMT quantum efficiency, and also highlight future work that is required to better understand the factors associated with the operational parameters of the  MCP-PMT, the properties of charge sharing, and the calibration of the readout electronics.

A half-sized LHCb demonstrator module with a $660 \times 1250 \times 10\,\rm{mm^3}$ radiator plate has been constructed and is currently being evaluated \cite{Hancock_2019}. The demonstrator has been instrumented with the same $8 \times 64$ MCP-PMT and readout electronics as for the June 2018  beam test, alongside a second identically-configured $8 \times 64$  MCP-PMT with an improved quantum efficiency. This will allow timing
resolution studies and photon-yield measurements with improved photon statistics.
Analysis is underway and will be the subject of a future paper.

\section*{Acknowledgements}
The support is acknowledged of the Science and Technology Research Council, UK, (grant number ST/P002692/1) and of the European Research Council through an FP7 Advanced Grant (ERC-2011-AdG 299175-TORCH).
The authors wish to express their gratitude to the CERN EP-DT-EF group, Simon Pyatt of the University of Birmingham, and to Gale Lockwood of the University of Oxford/RAL for their efforts in wire bonding the NINO chips. We also thank Andre Rummler for his support on the test-beam telescope.  We are grateful to Nigel Hay, Dominic Kent and Chris Slatter of Photek, and to Jon Lapington of the University of Leicester for their work on the MCP-PMT development. TB acknowledges support from the Royal Society, UK.



\bibliography{References}

\end{document}